\documentstyle[epsfig,prd,aps,floats]{revtex}   
\def\@normalsize{\@setsize\normalsize{10pt}\ixpt\@ixpt}

        \title{Constraints on Field Theoretical Models for Variation of the Fine Structure Constant}
	\author{Charles L. Steinhardt \\ \textsl{{Department of Astrophysical Sciences, Peyton Hall, Princeton, NJ 08540 \\ Center for Astrophysics, 60 Garden Street, Cambridge, MA 02138\thanks{From September 2003}}}}
	\begin{document}
\date{\today}		

\maketitle

\abstract{Recent theoretical ideas and observational claims suggest that the fine structure constant $\alpha$ may be variable.  We examine a spectrum of models in which $\alpha$ is a function of a scalar field.  Specifically, we consider three scenarios: oscillating $\alpha$, monotonic time variation of $\alpha$, and time-independent $\alpha$ that is spatially varying.  We examine the constraints imposed upon these theories by cosmological observations, particle detector experiments, and ``fifth force'' experiments.  These constraints are very strong on models involving oscillation, but cannot compete with bounds from the Oklo sub-nuclear reactor on models with monotonic time-like variation of $\alpha$.  One particular model with spatial variation is consistent with all current experimental and observational measurements, including those from two seemingly conflicting measurements of the fine structure constant using the many multiplet method on absorption lines.}}

\section{Introduction}
\label{sec:intro}

The conventional view is that the fine structure constant $\alpha$, the coefficient that determines the strength of the electromagnetic interaction, is a constant.  There is an effective change in $\alpha$ at energies greater than 1 GeV due to renormalization effects associated with electroweak symmetry breaking.  However, at temperatures far below the GeV scale, and certainly since matter-radiation equality, these effects are negligible.  

There are recent claims of a statistically significant variation of the fine structure constant at large redshifts $0.2 < z < 3.7$ (cf. \cite{Webb2003}),
\begin{equation}
\frac{\alpha(0.2 < z < 3.7) - \alpha(z = 0)}{\alpha(z = 0)} = (-0.57 \pm 0.10) \times 10^{-5},
\label{MWresult}
\end{equation}
where these measurements are averaged over quasar absorption systems in the given redshift range.  An independent measurement using similar techniques is reported in\cite{Chand2004},
\begin{equation}
\frac{\alpha(0.4 < z < 2.3) - \alpha(z = 0)}{\alpha(z = 0)} = (-0.06 \pm 0.06) \times 10^{-5}.
\end{equation}
One issue discussed in this paper is whether these two results are contradictory.  The only other possible statistically significant detection of variation of a coupling constant comes from measurements of isotopic abundances from the sub-nuclear reactor at the Oklo mine, where the analysis allows two possibilities for the value of $\alpha$\cite{Olive2002,Fujii2000,Fujii2002}.  The favored branch is a null result providing a strict bound on variation, but the other branch would also yield statistically significant variation. A full discussion of experimental bounds on the variation of coupling constants is given in \cite{Uzan2003}.

There is also considerable theoretical motivation for the possibility of variable coupling constants.  Coupling constants appear to be naturally variable in unified theories \cite{Cremmer1977}, and among other things will vary with a changing dilaton \cite{Damour1994}.  A changing fine structure constant would be accompanied by variation in other fundamental parameters, including the grand unification scale \cite{Fritzch2002}.  Variation in the fine structure constant also can be associated with violations of local Lorentz invariance and CPT symmetry \cite{Kostelecky2002}.  

In this paper, we consider experimental and observational constraints on field theories allowing variation of the fine structure constant, as well as whether any simple models satisfy both these constraints and the measurements of Webb et al.  In {\S~\ref{subsec:knownvar}}, we consider whether finite temperature field theory provides a sufficient correction to account for these observations.  In the remainder of the paper, we consider three classes of models and discuss phenomenological constraints that must be imposed.  Models with predominantly oscillatory behavior in the recent past are discussed in {\S~\ref{subsec:oscillate}}.  Models with monotonic variation are considered in {\S~\ref{subsec:time}}.  Finally, spatially varying but time-invariant models are considered in {\S~\ref{subsec:space}}.  For the remainder of this paper, we will use spatial variation to refer to models in which the value of $\alpha$ is different when measured at two points separated by space-like vectors and similarly time variation will refer to differences between points separated by time-like vectors.   While not every model will fall strictly into one of these three categories, the goal here is to produce representative constraints applicable to a wide range of models.

\section{General Considerations}
\label{sec:theory}

If a coupling constant is variable, we can treat it as a function of some field.  This paper will consider models in which the fine structure constant is a function of a scalar field.  Clearly not every possible scenario can be described in terms of a scalar field, but for small changes in $\alpha$ a scalar field will either be exact or a good approximation for a very wide variety of models.  Models that cannot be described in terms of a scalar field fall outside the scope of this paper.

The QED Lagrangian including a real scalar field $\phi$ is
\begin{equation}
{\cal L}_{QED} = \frac{1}{2}(\partial_\mu\phi)^2 - V(\phi) + \bar{l}\left(i\gamma_\mu\partial^\mu + e\gamma^\mu A_\mu - m_l\right)l - \frac{1}{4e^2}\left(\partial_\mu (eA_\nu) - \partial_\nu (eA_\mu)\right)^2,
\label{form1}
\end{equation}
where $l$ is summed over all leptons $l=e, \mu, \tau$, and $A_\mu$ is the vector potential.  Note that for constant $e$ and under the rescaling $eA_\mu \rightarrow A_\mu$, (\ref{form1}) becomes the more familiar
\begin{equation}
{\cal L}_{QED} = \frac{1}{2}(\partial_\mu\phi)^2 - V(\phi) + \bar{l}\left(i\gamma_\mu\partial^\mu + \gamma^\mu A_\mu - m_l\right)l - \frac{1}{4e^2}F_{\mu\nu}F^{\mu\nu}.
\label{form2}
\end{equation}
Both forms of the Lagrangian have been given simply to illustrate the proper transformation that leaves the two entirely equivalent.  The form given in (\ref{form1}) is given as the primary form for ease of calculation.  

If $\alpha$ has $\phi$-dependence, the QED Lagrangian will include a term
\begin{equation}
\label{Lag3.1}
{\cal L}_{QED} = \cdots + \bar{l}\sqrt{4\pi\alpha(\phi)}\gamma^\mu A_\mu l.
\end{equation}

The equation of motion for a scalar field $\phi$ in an expanding universe is 
\begin{equation}
\label{EofM}
\ddot{\phi} + 3H(t)\dot{\phi} = -V^\prime(\phi),
\end{equation}
where $V^\prime$ denotes the derivative with respect to $\phi$.  

As an illustrative example, consider the case in which $\phi$ has a minimum at $\phi_0 \equiv 0$, and the potential $V(\phi)$ is the harmonic potential 
\begin{equation}
V(\phi) = \frac{1}{2}m_\phi^2\left(\phi-\phi_0\right)^2.
\end{equation}
Substituting, the equation of motion becomes
\begin{equation}
\ddot{\phi} + 3H(t)\dot{\phi} + m_\phi^2(\phi-\phi_0) = 0.
\label{damping}
\end{equation}

This is the equation for a damped harmonic oscillator, with damping term $3H(t)$ (which has units of $\textrm{time}^{-1}$), and therefore admits the following characteristic solutions today:
\addtolength{\leftmargin}{1cm}
\addtolength{\rightmargin}{1cm}
\begin{enumerate}
\item{{\bf Large $\mathbf{m_\phi}$}.  If $m_\phi \gg H$, $\phi$, and therefore $\alpha$, will be rapidly oscillating with period $T \ll 3H_0$, where $H_0$ is the value of the Hubble parameter today.  The universe out to a redshift of 3.7, where variation of the fine structure constant has been reported in Webb et al., is observed over a time period of approximately $1/H_0$.  Therefore, over our range of observations, these models would predict that $\alpha$ would appear to be oscillating.  Models with these features are discussed in {\S~\ref{subsec:oscillate}}.}
\item{{\bf Intermediate $\mathbf{m_\phi}$}.  If $m_\phi \sim H$, $\phi$, and therefore $\alpha$, could be oscillating with period $T \gg 3H$.  In this case, we should observe the variation of $\alpha$ as monotonic rather than oscillating over our range of observations.  Also note that the damping term yields a characteristic decrease in $\dot{\alpha}$ of one e-folding per $1/3H_0$.  Models with monotonic time variation of $\alpha$ are discussed in {\S~\ref{subsec:time}}.}
\item{{\bf Small $\mathbf{m_\phi}$}.  If $m_\phi \ll H$, $\phi$, and therefore $\alpha$, will be frozen at its present value.  In this case, variation in $\alpha$ would only be possible if the value of $\phi$ frozen in at different regions of space were different.  Such models are discussed in {\S~\ref{subsec:space}}.}
\end{enumerate}
While not every model will contain a minimum or be well modeled using a harmonic expansion about the minimum, it is straightforward to adapt the constraints on these three generic cases to a particular model. 


\section{Running Coupling Constants}
\label{subsec:knownvar}

Before considering modifications to the QED Lagrangian in order to explain variation of the fine structure constant of the sort reported in Webb et al., we should consider whether the already known variation in coupling constants from predictions of quantum field theory might be responsible for this sort of deviation.  Using the fine structure constant as an example, renormalization tells us that at high energies, the value of $\alpha$ must be different than at low energies since there are loop calculations involved.  For one-loop diagrams, this ``running'' of $\alpha$ can be approximated as (e.g., \cite{Atchison1982}), for large momentum transfer $Q \gg m_e$,
\begin{equation}
\alpha(Q) = \frac{\frac{e^2}{4\pi}}{1-\frac{e^2}{12\pi^2}\textrm{ln}\left(\frac{Q}{m_e}\right)},
\end{equation}
where $m_e$ is the mass of the electron and we have taken $\hbar = c = k = 1$, or natural units.  
This running of $\alpha$ has indeed been observed experimentally, with Burkhardt and Pietrzyk\cite{Burkhardt1995} reporting 
\begin{equation}
\frac{1}{\alpha}\left(91.1884\textrm{ GeV}\right) \approx {128.89}
\end{equation}
from measurements at $Q$ corresponding to the mass of the $Z$ boson. 

For $Q \ll m_e$, the running of $\alpha$ can be approximated by\cite{Peskin} 
\begin{equation}
\frac{\Delta\alpha}{\alpha} \sim \left(\frac{Q}{m_e}\right)^2.
\end{equation}

The resulting variation in $\alpha$ depends on the energy scales involved.  To consider whether the variation would show up in current measurements of $\alpha$, we consider the Si~{\sc iv} absorption line doublets used in \cite{Murphy2001}, the most precise measurements that do not involve the potential difficulties of the many multiplet method.  Si~{\sc iv} absorption lines come from the warm ionized interstellar medium (WIM), where the most energetic processes take place typically around $10000\textrm{ K}$\cite{WIM}. 

Taking $Q \sim 10000\textrm{ K} \approx 0.86\textrm{ eV}$
\begin{equation}
\frac{\Delta\alpha}{\alpha} \sim 3 \times 10^{-12}.
\end{equation}

In addition to changes in $\alpha$ from renormalization group effects leading to a running coupling constant, there will also be finite-temperature corrections to any process involving a loop diagram.  As a result, the Lamb shift will be altered and therefore the atomic spectrum of silicon.  Depending on the temperature and on the atomic number of the atom involved, there are two corrections to atomic spectra that must be considered.  Walsh\cite{Walsh1971} considers a correction to the Lamb shift dominant at high temperatures and at low atomic number, concluding that
\begin{equation}
\Delta E \approx 300\frac{T^4}{Z^4(m_e\alpha)^3}.
\end{equation}
Using silicon ($Z = 14$) as an example, this correction would be
\begin{equation}
\Delta E \approx 2.6 \times 10^{-22}\textrm{ eV}.
\end{equation}
Barton\cite{Barton1972} considers a second correction, dominant at lower temperatures and higher atomic number over that calculated by Walsh, of approximately
\begin{equation}
\Delta E =\frac{\pi\alpha^3Z^2T^2}{8m_e},
\end{equation}
or again using the example of silicon,
\begin{equation}
\Delta E \approx 2.1 \times 10^{-14}\textrm{ eV}.
\end{equation}
In both cases note that we have assumed a temperature of $10000\textrm{ K}$.  Since only the most energetic processes in the WIM take place on this energy scale, this is an upper bound on $\Delta E$ but might be a significant overestimate.

We can also consider whether these effects could explain the many multiplet method results described in \cite{Webb2003}.  These also involve measurements of ions such as Mg~{\sc ii}, Fe~{\sc ii}, Al~{\sc ii}, Al~{\sc iii}, Zn~{\sc ii}, and Cr~{\sc ii} that occur in the WIM.  Using the iron ($Z=24$) line as an example, the correction to the Lamb shift is now
\begin{equation}
\Delta E \approx 4.8 \times 10^{-14}\textrm{ eV}
\end{equation}
and the correction directly to $\alpha$ is again a few parts in $10^{12}$.  Since the typical fine structure transitions in iron are of order $0.4\textrm{ eV}$, these corrections are also far too small to cause the variation claimed in \cite{Webb2003}.  

\section{Oscillation}
\label{subsec:oscillate}

As in {\S~\ref{sec:theory}}, the equation of motion (\ref{damping}) allows for three typical behaviors of a massive scalar field in an expanding universe.  For $m_\phi \gg H$, the scalar field will oscillate with an amplitude that is damped by the term $3H(t)\dot{\phi}$.  As an example, approximate the universe between a redshift of $z=4$ and the present as matter-dominated, so that $H(t) = \frac{2t_0}{3t}$.  This is a good approximation for $0.5 < z < 4$ while as dark energy comes to dominate the Hubble parameter tends toward a constant value.  Solving for the matter-dominated case, we find the amplitude drops off as $t^{-1}$.  Since the scale factor $a(t) \propto t^{2/3}$, the amplitude of oscillation in the recent past increases with higher redshift as
\begin{equation}
A(z) \propto {(1+z)}^\frac{-3}{2}
\end{equation}
in the regime in which $m_\phi \gg H$.

Assume the value of $\alpha$ today is oscillating and that therefore the present value is at some random phase of the periodic cycle.
The strongest current bound on $\dot{\alpha}$ is\cite{Marion2002}
\begin{equation}
\frac{\dot{\alpha}}{\alpha} = (0.2 \pm 7.0) \times 10^{-16} \textrm{yr}^{-1}.
\label{osc2}
\end{equation}
With 95\% confidence, $\dot{\alpha}$ chosen at a random point from oscillation with amplitude $A(z)$ and frequency $f$ will be 
\begin{equation}
\left|\dot{\alpha}\right| > 0.08 A(z=0)f.
\end{equation}
Therefore, at the $2\sigma$ level the atomic clock constraint is
\begin{equation}
\label{Afbound}
A(z=0)f < 9 \times 10^{-15}\textrm{yr}^{-1}\alpha.
\end{equation}
This bound does not apply to oscillations with $T < 10^{-8}\textrm{s} \sim 10^{-26}/H_0$, as atomic clocks cannot detect variation more rapid than the characteristic timescale.

Further, one also must explain why measurements of the fine structure constant using isotopic abundances from the sub-nuclear reactor at the Oklo mine \cite{Olive2002,Fujii2002,Naudet,Shlyakhter} yield a value so similar to the present value of $\alpha$.  The isotopic abundances found at the Oklo mine are consistent with a constant value of $\alpha$ that is sufficiently close to the present value as to reproduce the approximately 0.1 eV resonance for the interaction $^{149}_{62}\textrm{Sm} + n \rightarrow ^{150}_{62}\textrm{Sm} + \gamma$.  It has been shown that if $\alpha$ can be treated as constant during the lifetime of the reactor, the fine structure constant was within $1.44 \times 10^{-8}\alpha$ of the present value while the reactor was operating\cite{Fujii2002}.

For an oscillating model of $\alpha$ with sufficiently low frequency that the value of $\alpha$ can be treated as constant during the approximately $(2.3 \pm 0.7) \times 10^5\textrm{ yr}$ period of activity, the Oklo reactor could only operate at a time when the value of $\alpha$ were fixed so that the 0.1 eV resonance was reproduced.  Assuming that the frequency is sufficiently high that there have been a large number of periods in the intervening 1.8 billion years, a slight change in the frequency causes the present value of $\alpha$ to sweep through multiple periods, so that we should again expect to be at a random phase today.  As before, we require that at least 5\% of possible phases would produce a present value within the tolerances set by the Oklo measurements.  This is equivalent to a requirement at the $2\sigma$ level that
\begin{equation}
\label{Abound}
A(z=0) < 1.4 \times 10^{-7}\alpha.
\end{equation}
No work has been done that would rule out the possibility that in addition to being a good fit with a constant value of $\alpha$, the isotopic abundances at Oklo might also be a good fit for a model in which the value of $\alpha$ and therefore the energy of the neutron capture would vary in a specified manner during the reactor's operation.  

Therefore, for oscillation at an unknown frequency the results from the Oklo mine strongly suggest a limitation on the amplitude of oscillation.  We can further combine the atomic clock bound with observational constraints on frequency to produce an additional bound on the amplitude.  For example, Varshalovich et al\cite{Varshalovich1999} measures $\alpha$ from the spectra of 1500 absorption systems between $z=2$ and $z=4$ (or a time window $\Delta t \approx 0.15/H_0$), for an average sampling frequency of $f_s \sim 10000H_0$.  If there were oscillation within that time window with amplitude $A(z=3) > 4 \times 10^{-5} \alpha$, the oscillation would be detected in these results unless the frequency were greater than the Nyquist frequency, requiring
\begin{equation}
\label{fbound}
T > 0.15/H_0\textrm{ or }T < 1/{5000 H_0}.
\end{equation}
The low frequency case is really monotonic time variation, and will be considered in {\S~\ref{subsec:time}}.  Combining these limits with (\ref{Afbound}), the allowable frequency range requires $A < 3 \times 10^{-8} \alpha$, so we can conclude from a second set of constraints independent of Oklo that $A(z=3) < 4 \times 10^{-5} \alpha$, or equivalently
\begin{equation}
A(z=0) < 5 \times 10^{-6} \alpha.
\end{equation}

As another example, we can consider whether oscillation could be consistent with the results of Webb et al.  Clearly oscillation cannot be consistent simultaneously with both the variation claimed in Webb et al. and with the null result reported by Chand et al., but we will consider just the Webb results in isolation as an example of the difficulty of constructing an oscillatory model given present bounds.  In that case, the amplitude of the oscillation today must be at least 
\begin{equation}
A(z=0) \approx (0.81 \times 10^{-5})\alpha.
\end{equation}
This is larger than the bound from (\ref{Abound}) and the coincidence between the values at Oklo and today has a probability of 0.08\%.  The second (less likely) branch of Oklo would increase the probability to 1.6\%, still requiring greater than a 2$\sigma$ deviation from expectation.  

$A(z=0)$ is also larger than $5 \times 10^{-6} \alpha$, and therefore the Varshalovich bounds on the frequency will apply.  It is possible that future measurements will be sufficiently sensitive to evade the Varshalovich bounds on allowed frequencies.  In that case, one can analyze these new measurements in a similar manner to the analysis performed on the Varshalovich data in order to produce applicable bounds.  

Returning to the example of Webb et al., we can now calculate the probability that an oscillating model with the given $A(z=0)$ is consistent with atomic clock measurements.  Including the Varshalovich bounds on the frequency and assuming that $\alpha$ is at a random phase today, atomic clock limits on $\dot{\alpha}$ today will be satisfied with probability $1.1 \times 10^{-4}$.  For the claimed variation to be the result of an oscillating model, both this bound and the apparently independent bound from Oklo must be satisfied, for a combined probability of $9 \times 10^{-8}$.  

In general, we can see that the constraints on models in which the fine structure is oscillating are very strong in the absence of additional physics that would both finely tune atomic clock measurements today and explain why the sub-nuclear reactor at Oklo was active during a period when the value of the fine structure constant was the same as it is today.  At this point, an improvement of the precision with which $A$ can be constrained observationally would strengthen bounds on oscillating models.  However, additional observations of absorption systems at the present level of precision will only serve as additional confirmation that oscillation does not exist in a region already ruled out by current observational and experimental bounds.

\section{Monotonic Time Variation from Scalar Fields}
\label{subsec:time}

In order to produce an $\alpha$ that is monotonically varying we can introduce a scalar field $\phi$ and choose a function $\alpha(\phi)$ such that the QED Lagrangian is (writing only the electron terms)
\begin{equation}
{\cal L}_{QED} = (\partial_\mu\phi)^2 - V(\phi) + \bar{\psi_e}\left(i\gamma_\mu\partial^\mu + \sqrt{4\pi\alpha(\phi)}\gamma^\mu A_\mu - m_e\right)\psi_e - \frac{1}{4e^2}\left(\partial_\mu (eA_\nu) - \partial_\nu (eA_\mu)\right)^2.
\end{equation}
The equation of motion (\ref{EofM}) requires $\sqrt{V^{\prime\prime}} \approx 3H_0$, since small values of $V^{\prime\prime}$ result in a frozen-in field $\phi$ and large values result in oscillation with frequency $\gg 3H_0$.  

Any function $\alpha(\phi)$ is possible.  In general, we can take a Taylor series
\begin{equation}
\alpha \ {\equiv} \ \alpha({\phi})=\alpha_0+\lambda_1(\phi-\phi_0) + \frac{1}{2}\lambda_2\left(\phi-\phi_0\right)^2 + \cdots ,
\end{equation}
where $\lambda_n = \partial^n\alpha{}/\partial\phi^n$ and where we will take $\phi_0 \equiv 0$.  How could one test for such a $\phi$?  
Mass renormalization leads to a contribution to the masses of the proton and neutron that depends on $\alpha$\cite{Gasser1982}:
\begin{eqnarray}
\Delta m_p = B_p\Delta\alpha \ {\approx} \ (0.63 \textrm{ MeV})\frac{\Delta\alpha}{\alpha} \\
{\Delta}m_n = B_n\Delta\alpha \ {\approx} \ (-0.13 \textrm{ MeV})\frac{\Delta\alpha}{\alpha},
\end{eqnarray}
defining $B_p$ and $B_n$ coupling constants for the proton and neutron, respectively.

Dvali and Zaldarriaga\cite{Dvali2002} (see also Damour and Polyakov \cite{Damour1994}) consider the example $\alpha(\phi)$ = $\lambda \phi$, which leads to a Yukawa interaction 
\begin{equation}
{\cal L}={\cdots}+\sum_{N=p,n}\left[B_N\lambda_1\phi\bar{N}N\right].
\end{equation}
\begin{figure}[!ht]
\epsfxsize=1.5 in \centerline{\epsfbox{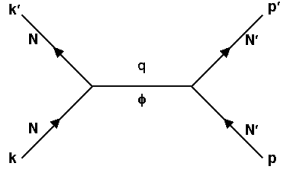}}
\caption{\label{fig:Yuklinear}Feynman Diagram for the Yukawa interaction between two nucleons $N$ and $N^\prime$ mediated by one $\phi$ particle.}
\end{figure}
They show this is a long-range force proportional to $\lambda_1^2$ (see Figure \ref{fig:Yuklinear}) but different for protons and neutrons, and that E\" otv\" os experiments require
\begin{equation}
\lambda_1^2 < \frac{10^{-10}}{M_P^2}.
\end{equation}

\begin{figure}[!ht]
\epsfxsize=1.5 in \centerline{\epsfbox{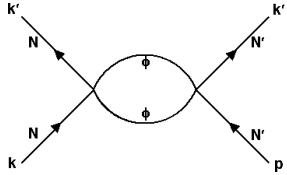}}
\caption{\label{fig:feynmanphis}Feynman Diagram for the $\phi^2$ interaction of two nucleons $N$ and $N^\prime$.}
\end{figure}

We can similarly consider the coupling $(\lambda_2/\Lambda)\phi^2\bar{N}N$.  As shown in Figure \ref{fig:feynmanphis}, $\lambda_2\phi^2\bar{N}N$ and by analogy the other higher-order terms in $\phi$ also produce long-range forces, mediated by $n$ $\phi$ particles rather than just one.  The $\Lambda$ is the cutoff energy from renormalization, and has been brought outside of $\lambda_2$ in order to keep all of the $\lambda_n$ dimensionless.

In comparing the two long-range forces, the matrix element corresponding to the interaction mediated by $n$ $\phi$ particles can be approximated in terms of the matrix element for the single $\phi$ interaction as 
\begin{equation}
M_n \approx A\left(\frac{\frac{\lambda_n}{\Lambda^{n-1}}}{\lambda_1}\right)^2M_1,
\end{equation}
since the matrix element is proportional to the coupling coefficient squared.  A is a constant with units of $\textrm{energy}^{n-1}$, and can be approximated as using the center of mass energy $\sqrt{s}$ as $\sqrt{s}^{n-1}$.  The cutoff energy $\Lambda$ is typically taken to be on the TeV scale.  For a long-range interaction between nucleons separated by a distance $r$, $\sqrt{s}$ typically scales as $1/r$, and therefore the long-range interaction between two nucleons will have $\sqrt{s} \ll \Lambda$.  The $\phi^n$ term produces a long range force with magnitude
\begin{equation}
F_n \approx \left(\frac{\lambda_n\sqrt{s}^{n-1}}{\lambda_1\Lambda^{n-1}}\right)^2F_1.
\end{equation}

The long range force is thus depressed by $\left(\frac{\sqrt{s}}{\Lambda}\right)^{2n-2}$.  Regardless of whether the force is mediated by one $\phi$ particle or many $\phi$ particles, the contributions from each force will simply add since the external lines on each diagram are identical.  Therefore, we can conclude from E\" otv\" os experiments that 
\begin{equation}
\left(\lambda_1 + \lambda_2 \left(\frac{\sqrt{s}}{\Lambda}\right) + \cdots\right)^2 = \left(\frac{\partial\alpha}{\partial\phi} + \frac{1}{2}\sqrt{s}\frac{\partial^2\alpha}{{\partial\phi}^2} + \cdots\right)^2 < \frac{10^{-10}}{M_P^2}.
\label{lblambda}
\end{equation}
%
This is effectively an upper bound on the coupling of $\alpha$ to $\phi$, although its utility depends on the model, as the higher-order derivatives of $\alpha(\phi)$ are only very weakly bounded.  We will calculate some examples shortly.

First, we must bound $\Delta\phi$ in order to use bounds on $\frac{\partial^n\alpha}{\partial\phi^n}$ to bound $\Delta\alpha$.  
For a flat universe, the kinetic energy density of our $\phi$ field cannot be greater than the critical density.  Dvali and Zaldarriaga\cite{Dvali2002} use this to show that $\lambda_1 > \frac{10^{-7}}{M_P}$.  However, this is only a bound on the present value.  Not only must the kinetic energy density fit this bound today, but it also must fit similar bounds in the distant past.  In an expanding universe, the energy density of a type of energy with equation of state parameter
\begin{equation}
\label{wdefined}
w = \frac{p}{\rho} = \frac{\frac{\dot{\phi}^2}{2} - V(\phi)}{\frac{\dot{\phi}^2}{2} + V(\phi)}
\end{equation}
changes as a function of scale factor $a$ as
\begin{equation}
\rho \propto a^{-3(1+w)},
\end{equation}
so that kinetic energy ($w = 1$) drops off as $a^{-6}$.  Therefore, if there is a little kinetic energy now, we should expect a lot of kinetic energy when the universe was smaller, or equivalently at large redshift ($a = 1/(1+z)$).  For example, at a redshift of $z=1$, the ratio of kinetic energy density to matter density would already be eight times larger than today; at a redshift of nine it will be 1000 times larger.  The effects of even a small amount of kinetic energy today would be enormous in the early universe, as at high redshift it would dominate the universe.  Among the effects we would have noticed would be a severe suppression of density fluctuations and a very large $\Delta\alpha$. 

This does not prove that $\phi$ cannot be rapidly changing today, but rather that, if $\phi$ is changing, in addition to the kinetic energy of the field there must be a sufficiently large potential that $w_\phi \lesssim 0$.  Much of the ``missing'' energy in the universe, or the so-called dark energy, has $w < 0$, so it might be natural to postulate that the scalar quintessence field postulated to solve the dark energy problem is also responsible for variation in $\alpha$. 

Having bounded $\Delta\phi$, we can now return to our bounds on $\frac{\partial^n\alpha}{\partial\phi^n}$ and use them to bound $\Delta\alpha$.  In order to highlight the limitations of the E\" otv\" os constraints, let us choose as a toy model an $\alpha(\phi)$ dominated by the $\phi^n$ term.  We choose
\begin{equation}
\alpha = \alpha_0 + K \phi^n,
\end{equation}
with $\phi = 0$ at the present value of $\alpha$.  The bound from (\ref{lblambda}) thus yields
\begin{equation}
\label{lblambdan}
\frac{\sqrt{s}}{n!}\frac{\partial^n\alpha}{\partial\phi^n} < \frac{10^{-5}}{M_P}.
\end{equation}
In this case, 
\begin{equation}
\Delta\alpha = K \left(\Delta\phi\right)^n = \frac{\partial^n\alpha}{\partial\phi^n}\frac{\left({\Delta\phi}\right)^n}{n!},
\end{equation}
so (\ref{lblambdan}) becomes
\begin{equation}
\frac{\Delta\alpha}{\alpha} < \frac{10^{-3}\left({\Delta\phi}\right)^n}{M_P\sqrt{s}^{n-1}}.
\end{equation}
Since $\Delta\phi$ can be as large as $M_P \gg \sqrt{s}$, the resulting bound of
\begin{equation}
\frac{\Delta\alpha}{\alpha} < 10^{-3}\left(\frac{M_P}{\sqrt{s}}\right)^{n-1}
\end{equation}
does not provide a useful constraint for $n>1$.

Dvali and Zaldarriaga consider the specific case $n=1$, for which $\frac{\Delta\alpha}{\alpha}$ may be as large as $10^{-3}$ between the present value and observations at a redshift of $2$.  However the bound from the Oklo reactor is much more restrictive.  For their model with constant $\dot{\phi}$ and thus constant $\dot{\alpha}$, at the time of Oklo we should measure
\begin{equation}
\left(\frac{\Delta\alpha}{\alpha}\right)_{\textrm{Oklo}} \sim 10^{-6}.
\end{equation}
Both branches of Oklo bounds rule out variation of this order.  Indeed, at the present time the Oklo bounds are the strictest bounds on monotonic time variation.

Bounds on violation of the equivalence principle have improved by a few orders of magnitude over the past decade or two\cite{Will2001}.  The Satellite Test of the Equivalence Principle (STEP) has been proposed to improve the sensitivity of E\" otv\" os experiments by five or six orders of magnitude\cite{STEP} and therefore the bounds on $\partial\alpha/\partial\phi$ may in the near future become more restrictive on a linear coupling than the measurements from Oklo.   Particle accelerator results such as those discussed in {\S~\ref{subsec:space}} will be orders of magnitude weaker even than the E\" otv\" os bounds.

One might be tempted to conclude that because E\" otv\" os experiments are only capable of bounding models with $\alpha = \alpha_0 + K \phi^n$ for $n=1$, models with $n>1$ are capable of producing large variation in $\alpha$ between $z=0$ and $z=1$.  Indeed, the best direct bounds on models with this type of time variation come from terrestrial bounds on and observations of the fine structure constant itself rather than from experimental bounds on associated effects such as violation of the equivalence principle.

However, in order to produce such a model, the linear term in $\alpha(\phi)$ must be very small and only the higher-order $\phi$-derivatives of $\alpha$ contribute.  In the absence of a symmetry (for example, a function of leading order $\phi^2$ would require the symmetry $\phi \rightarrow -\phi$), this is an unnatural model.  Even with a symmetry, the linear term will only disappear in an expansion about the minimum.  Since the E\" otv\" os experiments that constrain $\partial\alpha{}/\partial\phi$ take place today, this requires that the value of $\alpha$ be very near a minimum at present.  Since we are dealing with monotonic oscillation, the requirement is for the value of $\alpha$ to be very close to the minimum today but to be far from the minimum both in the recent past and in the near future.  Therefore just like the oscillating models discussed in {\S~\ref{subsec:oscillate}}, such a model requires a very improbable coincidence.

We do know that something fundamental has changed about the universe between a redshift of $3.5$ and today.  The universe has gone from decelerating to accelerating, and has gone from matter-dominated to a universe dominated by a dark energy with negative pressure.  Therefore, one might naturally consider a model in which either $\alpha$ or $\phi$ is a function of the composition of the universe, perhaps a non-minimally coupled scalar field (e.g., the Brans-Dicke\cite{BransDicke} model with a scalar field coupling to the Ricci scalar).

The coincidence between the timing of dark energy and of the reported variation in $\alpha$ from Webb et al. suggests that the two might be related.  Wetterich\cite{Wetterich2003} suggests that such a model might satisfy both variation on the order reported in Webb et al. and the Oklo bounds.  In order to produce the large variation between $z=0.1$ and $z=2$ compared to the variation between today and $z=0.1$, the dependence of $\alpha$ on dark energy must be similar to a step function.  

Assume that the dark energy is a time-varying quintessence field $Q$ rather than a cosmological constant.  Then the kind of dependence one is looking for might be of the form
\begin{equation}
\alpha(\Omega_Q) = \alpha_0 + \alpha_1e^{\frac{-\Omega_Q}{Q_0}}.
\end{equation}
Letting, as an example, $Q_0 = 0.1$, $\alpha$ would take on a constant value in the distant past as well as a constant value today, but the constant value would be different in those two epochs.  Similar ideas have been discussed by Barrow\cite{Barrow2002} among others.  However, no such model that fits both the variation reported by Webb et al. and all of the other measured bounds, from not only Oklo but also the CMB\cite{Martins2003} and nucleosynthesis\cite{Bergstrom1999}, has yet been worked out in detail.  While Barrow\cite{Barrow2003} suggests that a model with some features in common with the one proposed here will have the correct asymptotic features, it is not at all clear that the boundary conditions can be matched in such a way as to actually fit current measured limits.  

 To this point, a model that both is consistent with the claimed variation in \cite{Webb2003} and arises naturally from current theory has not been forthcoming.  Additionally, for time variation the recent results reported by Chand et al. are in direct conflict with those of Webb et al.  Any model in which the variation of $\alpha$ is time requires that independent measurements of $\alpha$ over the same range of redshift must produce the same result.  Therefore an additional requirement of any model that produces variation of the sort claimed by Webb et al. by introducing some $\alpha(Q)$ is that either the Chand et al. results are shown to be erroneous or $\Omega_Q$ is shown to have already taken on its present local value in some regions of the universe at a redshift of $2.3$, which would be inconsistent with inflation.

\section{Spatial Variation}
\label{subsec:space}

Another logical possibility for producing a different value of $\alpha$ at $z=1$ than at $z=0$ is that rather than changing over time, $\alpha$ undergoes spatial variation.  This is perhaps the simplest way of removing the Oklo constraint, as spatial variation could produce a different value of $\alpha$ at large distance (or, equivalently, at large red shift) while retaining the same constant value on Earth in the distant past.  Oscillating models will run into similar constraints as {\S~\ref{subsec:oscillate}}, although only the bound (\ref{Afbound}) will be pertinent.

In order for spatial variation to occur, there must be (at least) two values of $\alpha$ for which it will remain constant.  One possibility, which we will not consider here, is that a light field, such as that postulated by ``tracker'' quintessence models, is stuck as in \cite{trackerquint}.  For small $V^\prime$ and $V^{\prime\prime} \ll H^2$, $\dot{\phi}$ will be small.  Such models typically produce small $V^\prime$ and $V^{\prime\prime}$ by requiring the field to couple to almost nothing else; since $\alpha$ must be a function of our scalar field it will couple to almost everything.  

The other possibility is a relatively heavy field in which the potential $V(\alpha)$ has at least two local minima, corresponding to the values of $\alpha$ locally and at large distances.  In the early universe, the temperature is very high, and for total energy much greater than the potential, $\phi$ can take on any value.  As the universe expands and cools, at some point $\phi$ falls into one of these local minima and cannot get out of the potential well.  However, the local value determines which well $\phi$ will fall into, and thus different regions of the universe can randomly fall into different potential wells.  Therefore, it is possible that we are in one potential well with one value of $\alpha$ and much of the universe is in another potential well with a different value.

As in the previous section, we can equivalently consider a real scalar field $\phi$ such that $\alpha$ is a function of $\phi$.  We select the minima to be $\pm\sigma$.  We can then write the Lagrangian as
\begin{equation}
\label{DMLag}
{\cal L} = \frac{1}{2}\left(\partial_\mu\phi\right)^2 - \frac{1}{4}\lambda\left(\phi^2 - \sigma^2\right)^2, 
\end{equation}where $\lambda$ is typically of order unity.

At high temperatures, $\partial_\mu\phi \gg \phi^2 - \sigma^2$, and therefore the expectation value $\left<{\phi}\right> = 0$.  However, at lower temperatures $\phi$ will take on a vacuum expectation value of either $\left<\phi\right> = \sigma$ or $\left<{\phi}\right> = -\sigma$, spontaneously breaking the ${\cal Z}_2$ symmetry under $\phi \rightarrow -\phi$.  We might imagine that the local vacuum has $\left<\phi\right> = \sigma$ and that the vacuum containing absorption systems measured by Murphy et al. has $\left<\phi\right> = -\sigma$.  Since $\phi(x^\mu)$ must be continuous, even at very low energies there will be a region of space called a domain wall for which $\phi = 0$.  The major problem that we shall run into will be the total energy density of these walls.

A scalar field has stress tensor
\begin{equation}
\label{tmn}
T_{\mu\nu} = \partial_\mu\phi\partial_\nu\phi - g_{\mu\nu}{\cal L}.
\end{equation}
Because the domain wall is in a region of false vacuum, instead of $T_{\mu\nu} = 0$ the potential term of the Lagrangian yields \cite{KT} 
\begin{equation}
\label{DWT}
T^0{}_{0} = \frac{1}{2}\lambda\sigma^4{\textrm{cosh}}^{-4}\left(\lambda^{1/2}{\sigma}z\right)
\end{equation}
for a wall located in the $x-y$ plane centered at $z=0$.

So, what is the energy of the domain wall network?  For small values of $z$, i.e. values at and around the domain wall, 
\begin{equation}
T^0{}_{0} \approx \frac{1}{2}\lambda\sigma^4.
\end{equation}
As a lower bound on the wall energy density $\rho_w$, consider the case in which there is only one domain wall in the observable universe.  Then, $\rho_w$ will be the surface energy density multiplied by the wall thickness and divided by the horizon size $r$, or
\begin{equation}
\rho_w \approx \frac{1}{2}\lambda\sigma^4z_w/r.
\end{equation}
From (\ref{DWT}), 
\begin{equation}
z_w = \frac{1}{\lambda^{1/2}{\sigma}}.
\end{equation}
Note that we have set $\hbar = c = 1$.  For a flat universe (i.e. k=0) such as ours, 
\begin{equation}
r(t) = a(t)\int_0^t{\frac{dt}{a(t)}} \approx \frac{1}{H(t)}.
\end{equation}
Since we can expect this symmetry to have been broken during the radiation-dominated era (more on this shortly), the energy density of the universe is $\rho \approx T^4$, and by the Friedmann equation
\begin{equation}
H(t)^2 = \frac{8{\pi}\rho}{3M_P^2}.
\end{equation}
Further, the symmetry is broken when the thermal energy drops low enough that $\phi$ gets trapped in one of the minima of the double well potential in (\ref{DMLag}), or when $T \approx \sigma\lambda^{1/4}$. 

Therefore, the total energy density of the domain wall is
$$
\rho_w \approx \frac{1}{2}\frac{\lambda\sigma^4}{\lambda^{1/2}\sigma}\sqrt{\frac{8\pi}{3}}\frac{\sigma^2}{M_P}
$$
\begin{equation}
\rho_w \approx \frac{\lambda^{1/2}\sigma^5}{M_P}.
\end{equation}
Since the universe is flat, the critical density at that time is very nearly the radiation density 
$$\rho_c \approx \rho_r \approx T^4 \approx \lambda\sigma^4.$$
So, at the time of spontaneous symmetry breaking (SSB), the ratio of the energy density in domain walls to the critical density will be
\begin{equation}
\Omega_{w_{SSB}} \approx \frac{\sigma}{\lambda^{1/2}M_P}.
\end{equation}

Since $\alpha$ is an electroweak coupling constant, this symmetry cannot be broken until at or after electroweak decoupling at approximately 300 GeV\cite{KT}.  Since $M_P \approx 10^{19}$ GeV, for $\lambda \gg 10^{-31}$ (which is shown to be the case below), $\Omega_w \ll 1$ and therefore we can treat domain walls as negligible in the Friedmann equation describing the expanding universe.  WMAP bounds using the Integrated Sachs-Wolfe effect confirm that $\Omega_w < 10^{-4}$ today\cite{WMAPGeneral} compared to $\Omega_m = 0.28$, so domain walls can be ignored in the Friedmann equation for the entire history of the universe.

The energy density of a domain wall network is known to scale with the size of the universe as approximately $\rho_w \propto a^{-1.44}$\cite{VS}.  This will again underestimate $\rho_w$, since a solitary wall has $\rho_w \propto a^{-1}$ and the additional dropoff is due to coarsening, based on interactions between different causally connected walls.

Let the scale factor at the time of symmetry breaking be $a_w \equiv 1/(1+z_w)$,  where today $a_0 = 1$.  Then,
\begin{equation}
\Omega_{w_0} \approx \Omega_{w_{SSB}}\left(\frac{a_0}{a_w}\right)^{-1.44}\frac{\rho_{c_{SSB}}}{\rho_{c_{0}}}.
\end{equation} 
During the radiation-dominated era, i.e. $z \gg 3500$, $\rho_c \propto a^{-4}$ and during the matter-dominated era, $z \ll 3500$, $\rho_c \propto a^{-3}$.  (The effects of $\Lambda$-domination since a redshift of 0.5 or so are negligible).  So, for $1+z_w > 3500$, 
\begin{equation}
\label{spatiallb}
\Omega_{w_0} \approx \frac{\sigma}{\lambda^{1/2}M_P}(1+z_w)^{1.56}\frac{1+z_w}{3500}.
\end{equation}
If instead the symmetry is broken during the matter dominated era,
\begin{equation}
\Omega_{w_0} \approx \frac{\sigma}{\lambda^{1/2}M_P}(1+z_w)^{1.56},
\end{equation}
so (\ref{spatiallb}) is a lower bound on $\Omega_{w_0}$.

In general, the scale factor is inversely proportional to the temperature, $a \propto T^{-1}$.  We know that the symmetry is broken at approximately a temperature of $\sigma$.  Therefore, we can conclude that 
\begin{equation}
1+z_w \approx \frac{\sigma}{T_{CMB}},
\end{equation}
where $T_{CMB}$ is the current temperature of the universe, 0.000235 eV.  Substituting, we conclude that
\begin{equation}
\Omega_{w_0} \geq \frac{\sigma^{3.56}}{(5.65 \times 10^{-7})\lambda^{1/2}M_P},
\end{equation}
with $\sigma$ in eV.  

$\Omega_w < 10^{-4}$ today\nocite{WMAPGeneral}, so substituting we conclude that
\begin{equation}
\sigma \lambda^{-0.140} < 103 \textrm{ keV}.
\label{DWbound}
\end{equation}

Is this upper bound valid in all cases?  If the formation of walls occurs prior to inflation, the walls can be inflated far away, similar to the solution of the monopole problem.  It might then simply be an accident that one happens to lie in our observable universe.  However, as mentioned above, the electroweak decoupling only occurs around 300 GeV, well after inflation.  Further, if the domain wall were inflated out of our observable universe, our observable universe would all be in the same domain.  Therefore, this model could not produce variation in the fine structure constant.

Another way to prevent domain wall problems is restore the symmetry at a lower temperature.  At first, this might also seem problematic, since we want the spatial variation laid down initially to remain today.  Indeed, such a model does not properly fit under the spatial variation heading at the start of this section.  Due to causality, we cannot observe the present value of $\alpha$ at large distances, and therefore we cannot rule out the possibility that the value of $\alpha$ is different at present in gas clouds large distances than it was when light we observe today was passing through those regions.  However, since the observations of Webb et al. include systems at $z=0.5$, WMAP bounds on $\Omega_w$ using the Integrated Sachs-Wolfe effect since $z=2$ will continue to constrain such models.  

A third option would be for there to be a slight asymmetry, such that the two local minima are unequal.  In that case, one vacuum would be preferred with regions of higher-density vacuum shrinking and eventually becoming black holes.  However, this effect only saves the model from becoming wall-dominated when these regions start to disappear, so a model that postulates two regions of different vacuum today or in the recent past is incompatible.  

By introducing a scalar field $\phi$, we have also introduced a particle $\phi$.  What is the mass of this particle?  As in (\ref{DMLag}), 
\begin{equation}
{\cal L} = \frac{1}{2}\left(\partial_\mu\phi\right)^2 - \frac{1}{4}\lambda\left(\phi^2 - \sigma^2\right)^2. 
\end{equation}
The mass at the present value $\phi = \sigma$ will be $m_\phi = \sqrt{V^{\prime\prime}(\sigma)}$, or
\begin{equation}
m_\phi = \lambda^{\frac{1}{2}}\sigma\sqrt{2}.
\end{equation}
(As an aside, notice that had we allowed a complex scalar field it would also yield a particle of mass $\lambda\sigma\sqrt{2}$ along with a Goldstone boson.)

We can constrain $\lambda \sigma\sqrt{2}$ for a model of the sort that might fit measurable variation by requiring that the symmetry is broken after electroweak decoupling and before $z=3$, or
\begin{equation}
300 \textrm{ GeV} > T \approx \sigma\lambda^{1/4} > 10 \textrm{ K}.
\end{equation}
Combining this with the WMAP bound on domain wall density from (\ref{DWbound}), we find
\begin{equation}
\lambda^{\frac{1}{2}}\sigma\sqrt{2} < 5.7 \times 10^{-9} \textrm{ eV}.
\end{equation}

Because there is no lower bound on the wall density, there is no direct upper bound on $\lambda$.  As $\lambda$ gets large, constraints on $\sigma\lambda^{1/4}$ require that $\sigma$ approaches zero and the Lagrangian simply describes $\lambda\phi^4$ theory.  Having fixed $\lambda$ on some energy scale $M$, the effective $\bar{\lambda}$ will be\cite{Peskin}
\begin{equation}
\bar{\lambda}(p) = \frac{\lambda}{1-(3\lambda/16\pi^2)\log(p/M)}.
\end{equation}
When $(3\lambda/16\pi^2)\log(p/M) = 1$, the perturbation theory breaks down.  We require that the theory be valid for $p < 300 \textrm{ GeV}$, the range on which accelerator experiments may be able to test it.  If we fix $\lambda$ at $M = 91 \textrm{ GeV}$, the energy scale of the relevant L3 measurements discussed below, this requires $\lambda < 41.6$, and therefore
\begin{equation}
m_\phi < 760 \textrm{ GeV}.
\end{equation}

Because variation in $\alpha$ depends on the expectation value of $\phi$ rather than on the presence of the associated particle, it is acceptable for the mass of the particle to be far above the background temperature at the time of spontaneous symmetry breaking.  However, since in the near future there will be experimental results at energies greater than even the largest possible $m_\phi$, let us consider the case of a particle with mass less than 91 GeV and ask whether it would have been observed.

The QED Lagrangian includes a term
\begin{equation}
{\cal L}_{QED} = \bar{e}\sqrt{4\pi\alpha}\gamma^\mu A_\mu e,
\end{equation}
where $e$ is the electron field.  Since we have postulated a $\phi \rightarrow -\phi$ symmetry, we can expand $\alpha = \alpha_0 + \frac{\partial\alpha^2}{\partial\phi^2}$.  Since $\Delta\alpha = \alpha_1 - \alpha_0$ is small compared to $\alpha_0$, we can approximate $\sqrt{\alpha} \approx \sqrt{\alpha_0} + \frac{1}{2}\frac{\partial\alpha^2}{\partial\phi^2}$.  Over a change in $\phi$ of $2\sigma$ $\alpha$ changes by $\Delta\alpha$.  While we do not know the function $\alpha(\phi)$, we can estimate the resulting bound by assuming a constant $\frac{\partial\alpha}{\partial\phi}$.  Then, 
\begin{equation}
\frac{\partial\alpha^2}{\partial\phi^2} \approx \frac{\Delta(\alpha^2)}{\Delta(\phi^2)} = \frac{\Delta\alpha}{2\sigma^2}.
\end{equation}

So, the coupling $\bar{e}A_\mu{e}$ splits into two branches,
\begin{equation}
{\cal L} = \cdots + \sqrt{4\pi}\bar{e}\left(1 + \frac{1}{2}\left(\frac{\Delta\alpha}{2\sigma^2}\right)\phi^2\right)\gamma^\mu A_\mu e.
\end{equation}
Known electron-positron decays begin with $e^+e^-$ decaying first to a virtual photon, and then to real products.  We have now introduced a different decay in which $e^+e^-\rightarrow\gamma\phi\phi$.  No $\phi$ particles have been detected, so this decay should appear to be a single-photon decay, forbidden by conservation of momentum in the center of mass frame.  However, the coupling producing $\phi$ particles is a five-point interaction with mass dimension six, and is therefore Planck suppressed.  Therefore, even the best L3 bound\cite{L3} of $\gamma_{e^+e^-\rightarrow\gamma\phi\phi}/\gamma_{\textrm{other}} \leq 1.1 \times 10^{-6}$ cannot be used to constrain this theory.  

It is also possible that the ${\cal Z}_2$ symmetry in the effective Lagrangian is explicitly broken instead of spontaneously broken, and that there is a coupling $\bar{e}\phi A_\mu e$.  This would yield two unequal minima and the corresponding problems discussed earlier.  We can repeat the calculation above using a coupling to a single $\phi$, but even so the mass dimension will be five and the four-point coupling will be Planck suppressed.  

So, this model is consistent with every observational and experimental bound aside from the many multiplet results.  Webb et al. report 
\begin{equation}
\frac{\alpha(0.2 < z < 3.7) - \alpha(z = 0)}{\alpha(z = 0)} = (-0.57 \pm 0.10) \times 10^{-5},
\end{equation}
while Chand et al. report
\begin{equation}
\frac{\alpha(0.4 < z < 2.3) - \alpha(z = 0)}{\alpha(z = 0)} = (-0.06 \pm 0.06) \times 10^{-5}.
\end{equation}
The two measurements are inconsistent with one another if $\alpha$ is spatially uniform, but closer examination is required if $\alpha$ is spatially varying.  In particular, one might examine whether the discrepancy can be explained by the fact that the two groups examined different regions of the sky.  The Chand et al. results are based on a sample from the Southern Hemisphere only, whereas Webb et al. consider systems in both the Southern and Northern Hemisphere.   For simplicity, we divide the Webb et al. data into Southern only and Northern only.  (A more sophisticated study might break the Webb et al. dataset along a different  boundary, but our choice will suffice for the purposes of illustration.)  For 96 quasars in the Northern sample,
\begin{equation}
\left(\frac{\alpha(0.2 < z < 3.7) - \alpha(z = 0)}{\alpha(z = 0)}\right)_{\textrm{North}} = (-0.66 \pm 0.12) \times 10^{-5},
\end{equation}
while for 32 quasars in the Southern sample,
\begin{equation}
\left(\frac{\alpha(0.2 < z < 3.7) - \alpha(z = 0)}{\alpha(z = 0)}\right)_{\textrm{South}} = (-0.36 \pm 0.19) \times 10^{-5}.
\end{equation}

Therefore, this model allows for the possibility that both Chand et al. and Webb et al. are consistent despite appearance to the contrary.  Additional observations by both groups in both Hemispheres would allow a determination of whether the two sets of analysis agree on the value of $\alpha$ when looking at an identical patch of sky.  If so, and if both groups agree that there is a different value in parts of the Northern and Southern hemispheres, this would be strong evidence for spatial variation.  Such a test could be accomplished with an additional sample of approximately 50 quasars and would be capable of either ruling out or confirming whether such a model can explain the discrepancy.  Alternatively, this test might instead show a systematic difference between the results of the two groups in identical patches of sky, which would suggest that the apparent discrepancy should be resolved through a re-examination of systematic errors rather than through a new theoretical model.

\section{Discussion}
\label{sec:conclusion}

Observations from distant absorption systems suggest that the fine structure constant may not be constant, but rather varying as a function of distance or time.  While other measurements suggest the value of $\alpha$ might instead be constant, in this paper we have considered features generic to models that predict spatial or time variation.  In particular, we have considered models in which $\alpha$ varies as a function of some scalar field, since such models could be motivated either by string theory or more recently by cosmological attempts to solve the dark energy problem.  Scalar fields can cause any of three classes of variation of $\alpha$, considered in {\S~\ref{sec:theory}}.  

\begin{itemize}
\item{{\bf Oscillating $\alpha$.}  If $\alpha$ is oscillating rapidly in the recent past, the characteristic change in $\alpha$ is constrained by measurements of $\dot{\alpha}$ from atomic clocks today.  For oscillation of amplitude $A$ and frequency $f$, this requires 
\begin{equation}
A(z=0)f < 9 \times 10^{-15}\alpha
\end{equation}
except at frequencies greater than 100 MHz.  For sufficiently large amplitude, it would also be a coincidence that the value of $\alpha$ at Oklo is very nearly the same as $\alpha$ today.  This coincidence requires
\begin{equation}
A(z=0) < 1.4 \times 10^{-7}\alpha.
\end{equation}
}
\item{{\bf Monotonic time variation of $\alpha$.}  If $\alpha$ is changing monotonically in the recent past, the strongest bound comes from the Oklo sub-nuclear reactor.  For the simplest model with $\alpha(\phi) = \alpha_0 + K\phi$ and with constant $\dot{\phi}$ than the Oklo bound
\begin{equation}
\left|\dot{\alpha}\right| < 10^{-17}\alpha\textrm{ yr}^{-1}
\end{equation}
is stronger than any phenomenological bounds.  Projected E\" otv\" os-type experiments such as STEP may improve this bound.}

E\" otv\" os bounds on higher-order couplings of $\alpha$ to $\phi$ do not provide useful constraints.

\item{{\bf Spatial variation of $\alpha$.}  If $\alpha$ is changing with distance rather than with time, there are two possibilities.  A light, minimally coupled field could result in a monotonic $\alpha(x^i)$, but it is difficult to have a minimally coupled field that also causes a change in the ubiquitous fine structure constant.  Or, a heavier field with a discrete symmetry can undergo spontaneous symmetry breaking, resulting in different values of $\alpha$ in different regions of space.}
\end{itemize}

We have shown that the Chand et al. and Webb et al. results agree in the Southern Hemisphere, where their sampled regions overlap.  Both are consistent with no variation in $\alpha$ from the terrestrial value.  However, Webb et al. also include a sample from the Northern Hemisphere which does show a statistically significant deviation from the terrestrial value.  If $\alpha$ is uniform, then the two results are inconsistent.  However, we have found that a model in which $\alpha$ has different values in different domains (with one value for the Earth and Southern Hemisphere and another for the Northern Hemisphere) is consistent with both measurements and with all other current constraints.  Alternatively, the observed difference between Southern and Northern Hemisphere may be due to a systematic bias that has not yet been identified.  Further observations should clear up the situation.

Another possible explanation for spatial variation of coupling constants is the recent ``Chameleon'' model in which a scalar field can have a mass dependent upon the local matter density\cite{Khoury2003}.  Such a model would allow the $\phi$ field to take on one value terrestrially and another value in gas clouds, and therefore allows variation that cannot be constrained by Oklo or atomic clocks.  However, unlike the spatial variation case discussed in {\S~\ref{subsec:space}}, there is no artifact such as a domain wall that would restrict such a model.   The only additional constraint that may prove useful would be from nucleosynthesis, but there are no restrictions on the form of $\alpha(\phi)$, and therefore it should be possible to satisfy this constraint.  Similar behavior also may be obtained without a direct coupling to the density, as proposed by Barrow and Mota \cite{Barrow2003,Mota2003}.  To first approximation, gas clouds in different regions of the sky should have the same density, and therefore we should not observe spatial variation of $\alpha$ between gas clouds in different patches of sky.  However, since the desired effect on $\alpha$ is quite small, it maybe that only a tiny but systematic difference between the densities of gas clouds in one region of sky and in another is required to produce differences in $\alpha$ of a few parts per million.  Experiments scheduled for the next decade or two should be sufficiently sensitive to test predicted variation of the gravitational constant $G$ according to this model\cite{Khoury2003}.

Certainly if spatial variation of $\alpha$ were confirmed in the near future, this would provide an immediate and strong constraint on allowable models for unified theories.  However, it should be stressed that any scalar field introduced that would cause the variation of $\alpha$ is subject to the bounds discussed in this paper.  Therefore, improving current constraints on the variation of the fine structure constant and other coupling constants will continue to restrict particular models for unification.  As there are at present very few measurements of observations with the potential to restrict such theories, it is even that much more critical to resolve the issues with the two many multiplet results and to continue to improve these bounds.

\acknowledgements
The author also thanks Nima Arkani-Hamed, Justin Khoury, David Mota, and Herman Verlinde for helpful comments, as well as David Spergel for his guidance throughout this project.  This material is based in part upon work supported under a National Science Foundation Graduate Research Fellowship.

\end{document}